# Self-Optimized Superconductivity Attainable by Interlayer Phase Separation at Cuprate Interfaces

Takahiro Misawa[1*], Yusuke Nomura[2], Silke Biermann[2], Masatoshi Imada[1**]

[1]Department of Applied Physics, University of Tokyo,

7-3-1 Hongo, Bunkyo-ku, Tokyo, 113-8656, Japan,

[2]Centre de Physique Théorique, École Polytechnique,

CNRS, Université Paris-Saclay, F-91128 Palaiseau, France

[*]Present address: Institute for Solid State Physics, University of Tokyo, 5-1-5 Kashiwa-noha, Kashiwa, Chiba 277-8581, Japan.

[**]To whom correspondence should be addressed; E-mail: imada@ap.t.u-tokyo.ac.jp.

**Stabilizing superconductivity at high temperatures and elucidating its mechanism have long been major challenges of materials research in condensed matter physics. Meanwhile, recent progress in nanostructuring offers unprecedented possibilities for designing novel functionalities. Above all, thin films of cuprate and iron-based high-temperature superconductors exhibit remarkably better superconducting characteristics (e.g. higher critical temperatures) than in the bulk, but the underlying mechanism is still not understood. Solving microscopic models suitable for cuprates, here, we demonstrate that at an interface between a Mott insulator and an overdoped non-superconducting metal, the superconducting amplitude stays always pinned at the optimum achieved in the bulk, independently of the carrier concentration in the**

**metal. This is in contrast to the dome-like dependence in bulk superconductors, but consistent with the astonishing independence of the critical temperature from the carrier density $x$ observed at interfaces of $La_2CuO_4$ and $La_{2-x}Sr_xCuO_4$. We furthermore identify a self-organization mechanism as responsible for the pinning at the optimum amplitude: An emergent electronic structure induced by interlayer phase separation eludes bulk phase separation and inhomogeneities that would kill superconductivity in the bulk. Thus, interfaces provide an ideal tool to enhance and stabilize superconductivity. This interfacial example opens up further ways of shaping superconductivity by suppressing competing instabilities, with direct perspectives for designing devices.**

**Teaser: How and why can maximized superconducting amplitudes be self-organized at interfaces? The answer opens perspectives for design.**

**INTRODUCTION**

Thin films and interfaces offer unique platforms for designing materials functions, beyond what is possible in the bulk. Above all, superconductivity at interfaces was observed even in cases where the bulk compounds sandwiching the interface are both non superconducting. Furthermore, the critical temperatures $T_c$ in thin films higher or equal to the maximum $T_c$ of the bulk material were observed (*1, 2*). These findings suggest a superiority of interfaces for designing high $T_c$ superconductivity.

Our understanding of the cuprate high-$T_c$ superconductors in the bulk has not yet reached consensus and our knowledge stemming from experimental measurements has constantly been updated since the discovery of superconductivity in the cuprates. However, the dome structure of $T_c$ as a function of carrier (doping) concentration is a common property irrespective of the compounds. In particular, the optimum $T_c$ is realized only at a specific doping concentration δ around 0.15 per Cu.

In contrast, at the interface of $La_2CuO_4$ and $La_{2-x}Sr_xCuO_4$ (schematically illustrated in Fig.1 (A)), for $x$ in the range of 0.2 to 0.5, stable superconductivity with $T_c \sim 40K$ irrespective of the value of $x$ was observed (*2*). This discovery was even more unexpected as the value of this "pinned" $T_c$ is very close to the maximum value for the bulk superconductivity in $La_{2-x}Sr_xCuO_4$, realized there only for the optimum carrier doping concentration of $\delta = x \sim 0.15$. If the mechanism of this superiority and stability at the interface is understood, we may gain insight not only into the unsolved mechanism of superconductivity, but also into how to reach higher critical temperatures in elaborately designed devices.

For the bulk superconductivity of the cuprates, noteworthy theoretical progress was made recently: Numerical calculations using various tools have been able to reproduce the basic experimental characteristics, in particular the *d*-wave symmetry of the gap and the dome structure (*3–11*).

Using cutting-edge variational Monte Carlo simulations (*10, 12*) for a stacked layer model shown in Fig. 1(B) (top panel), here we show that superconductivity emerges dominantly at a single layer of the interface between a Mott insulator and an overdoped metal and its amplitude is independent of the carrier density in the metallic side. The amplitude is indeed pinned at the maximum of the dome structure in the bulk in perfect agreement with the experiment.

Our numerical result shows that this pinning originates from the underlying electronic phase separation in the bulk (*10, 13–15*), which by itself would destroy the superconductivity in the bulk but is now replaced by an interlayer phase separation around the interface instead of the phase separation within a layer (schematically shown in the bottom panels in Fig. 1(B)). In general, strong coupling superconductivity with a high critical temperature would require a strong effective attraction between electrons. However, this strong attraction works as a "double-edged sword". Namely, it also makes the system prone to charge inhomogeneity that destroys superconductivity. The interface cleverly eludes this trade-off.

## RESULTS

**Theoretical model**

We analyze the multi-layer single-band Hubbard model, which is suitable for studying interfaces of the cuprates, defined by

$$H = -t \sum_{\langle i,j \rangle} \left(c^\dagger_{i\sigma\nu} c_{j\sigma\nu} + \text{h.c.}\right) - t_z \sum_{i,\sigma,\langle \nu,\nu' \rangle} \left(c^\dagger_{i\sigma\nu} c_{i\sigma\nu'} + \text{h.c.}\right)$$

$$+ U \sum_{i,\nu} n_{i\uparrow\nu} n_{i\downarrow\nu} - \sum_{i,\sigma,\nu} \varepsilon_\nu n_{i\sigma\nu} \quad (1)$$

where $c^\dagger_{i\sigma\nu}(c_{i\sigma\nu})$ is the creation (annihilation) operator of an electron at $i$th site on the $\nu$th layer with spin $\sigma$ and $n_{i\sigma\nu} = c^\dagger_{i\sigma\nu} c_{i\sigma\nu}$ is the corresponding number operator. For simplicity, we consider only the nearest-neighbor pair for the intra-layer transfer $t$. For the inter-layer transfer we take $t_z = -0.05t$ and the onsite Coulomb interaction is set to $U = 8t$. These are realistic values in terms of first-principles and numerical estimates (*10, 16, 17*) compared to the experimental optical gap and transport measurements (*18, 19*). Hereafter, we set the energy unit $t = 1$ (~ 0.5 eV in the cuprates). The layer-dependent onsite hole level is represented by $\varepsilon_\nu$. We confirmed that details of the parameter values do not alter our results.

We perform high-accuracy many-variable variational Monte Carlo (mVMC) calculations at temperature $T = 0$ for an $N_s = L \times L$ square lattice stacked as a slab with thickness $L_{\text{layer}}$. (See Materials and methods section for details of the model and method.) The mVMC method is capable of describing quantum and spatial fluctuations (*10, 12*), allowing for an accurate estimate of the superconducting stability among the competing orders.

The experimental interface illustrated in Fig. 1(A) has a transient region caused by the inter-layer diffusion and exchange between La and Sr atoms (*2, 20*) (blue line in Fig. 1(C)). To realistically mimic the inter-layer diffusion effect that makes the onsite energy level gradually change within a few layer, we construct a slab around the interface, with the layer-dependent onsite level as $\varepsilon_{v+1} = \varepsilon_v - \Delta\varepsilon$ ($3 \geq v \geq 1$) with a constant $\Delta\varepsilon$ (see red line in Fig. 1(C)). The 0th layer is assumed to be insulating and the other layers ($v \geq 1$) become metallic. For the 0th layer, we employ $\varepsilon_0 = \varepsilon_1 + 1$, ensuring the insulating nature.

On the other hand, density functional theory calculations for a sharp interface predict a more abrupt change in the onsite energy level (see Fig. 1(D) and the Supplementary Materials for the first principles estimate.)

Note that the properties around the interface embedded in a sufficiently thick slab with $\varepsilon_v = \varepsilon_4$ for $v \geq 4$ can be well simulated by a slab of total thickness $L_z = 5$. This is because the transient region near the interface is confined to the region $v < 4$ if the Sr concentration is such as illustrated in Fig. 1(C). The density at $v \geq 4$ converges to a constant corresponding to the bulk value in the overdoped metallic side. In practical calculations, the bulk hole densities at $v = 4$ are controlled by changing $\Delta\varepsilon$ and the total electron number in the slab.

**Pinning of electron density at interface**

In Fig. 2(A), we plot the layer dependence of the hole density $\delta_v$ defined by $\delta_v = 1 - N_v/(L \times L)$, where $N_v$ is the average electron number in the $v$th layer (see Materials and methods section for the method used to determine the charge profile). The bulk hole density, $\delta_{bulk} = \delta_4$ monotonically increases with increasing $\Delta\varepsilon$. Experimentally, this corresponds to $x$ in the metallic side of the interface. Even if $\delta_{bulk}$ changes substantially, at the interface, $\delta_1$ is pinned.

To understand this counterintuitive pinning, we show the calculation for $\mu$-$\delta$ relation of a single layer in Fig.2(B), where $\mu$ is the chemical potential (*10*). We find essentially the same $\mu$-$\delta$ relation for the uniform bulk ($\mu=\mu_{bulk}$, $\delta=\delta_{bulk}$) consisting of stacked layers with the same single-particle level coupled by the small interlayer transfer $t_z = 0.05t$. Non-monotonic $\delta$ depedence of $\mu$ leads to a thermodynamic instability with the phase separation for $0 < \delta = \delta_{bulk} \lesssim \delta_{PS} \sim 0.20$ (see also Fig.4). The $\varepsilon_4$-$\delta_4$ relation traces the $\mu$-$\delta$ relation by the mapping $\varepsilon_4 \leftrightarrow 4\Delta\varepsilon\mu$ and $\delta_4 \leftrightarrow \delta$ (green curve in Figure 2(A)).

Remarkably, the $\mu_\nu$-$\delta_\nu$ and $\varepsilon_\nu$-$\delta_\nu$ relations at all the νth layers trace the same relation (see Fig. 4(B) and (C)).

This indicates that each layer is well represented by the single layer model and the effect of $t_z$ (=0.05$t$) is small as for the μ-δ relation. The main role of $t_z$ is to distribute the holes across the layers, where neighboring layers simply work as carrier reservoirs. A doping concentration that would lie in the region $0 < \delta < \delta_{PS}$ is prohibited at any layer. Consequently, the first layer that would lie in the phase separation region is in reality pinned at the border $\delta_{PS}$ as in Fig. 2(C). The consequences of the pinning of $\delta_1$ at $\delta_{PS}$ are further discussed later.

**Pinned superconducting order at interface**

To investigate the superconducting properties, we calculate the layer-dependent equal-time superconducting correlations of $d_{x^2-y^2}$-wave symmetry defined as

$$P_{\nu,d}(\boldsymbol{r}) = \frac{1}{2N_s} \sum_{\boldsymbol{r}_i} \left( \langle \Delta^\dagger_{\nu,d}(\boldsymbol{r}_i) \Delta_{\nu,d}(\boldsymbol{r}_i + \boldsymbol{r}) \rangle + \langle \Delta_{\nu,d}(\boldsymbol{r}_i) \Delta^\dagger_{\nu,d}(\boldsymbol{r}_i + \boldsymbol{r}) \rangle \right), \quad (2)$$

$$\Delta_{\nu,d}(\boldsymbol{r}_i) = \frac{1}{\sqrt{2}} \sum_j f_d(\boldsymbol{r}_j - \boldsymbol{r}_i)(c_{i\uparrow\nu} c_{j\downarrow\nu} - c_{i\downarrow\nu} c_{j\uparrow\nu}),$$

$$f_d(\boldsymbol{r}) = \delta_{r_y,0}(\delta_{r_x,1} + \delta_{r_x,-1}) - \delta_{r_x,0}(\delta_{r_y,1} + \delta_{r_y,-1})$$

where $\Delta_{\nu,d}$ denotes the $d_{x^2-y^2}$-wave superconducting order parameter at the νth layer, $f_d(\boldsymbol{r})$ is the form factor that describes the $d_{x^2-y^2}$-wave symmetry, and $\delta_{i,j}$ denotes the Kronecker's delta and $\boldsymbol{r} = (r_x, r_y)$ being the two-dimensional lattice coordinate scaled by the lattice constants of the square lattice.

In Fig. 3(A), we plot $P_{\nu,d}(\boldsymbol{r})$ for ν = 1 at $\Delta\varepsilon$ = 0.2 (blue squares). The superconducting correlation becomes a nonzero constant at the long-ranged part (essentially for $r = |\boldsymbol{r}| = \sqrt{r_x^2 + r_y^2} \geq 3$) at the interface layer (ν = 1) implying long-ranged order. Here, $P_{\nu,d}(\boldsymbol{r})$ is similar to the value for the uniform bulk system with a hole density similar to that at the interface. $P_{v,d}(\boldsymbol{r})$ for bulk system (red circles) is calculated for uniformly stacked layers that have the same single-particle level for all the layers (see Materials and Methods).

In Fig. 3(B), we plot the $\delta_{\text{bulk}}$ (metallic bulk density) dependence of the superconducting correlations in the long-range limit, which is, in practice, calculated from

$$\overline{P}_{v,d} = \frac{1}{M} \sum_{2 < r = |\boldsymbol{r}| < \sqrt{2}L} P_{v,d}(\boldsymbol{r}) \qquad (3)$$

for sufficiently large $L$, where $M$ is the number of vectors satisfying $2 < r < \sqrt{2}L$. In previous work (*10*), this quantity was shown to allow for a practical estimate of the long-range order and Fig. 3(A) also supports this criterion. Note that $\overline{P}_{v,d}$ converges to the square of the order parameter (superconducting amplitude) $\langle \Delta^\dagger_{v,d}(r) \rangle$ in the thermodynamic limit. We find in Fig. 3(B) that the (squared superconducting order parameter) is pinned irrespective of the bulk hole densities, in accordance with the pinned $\delta_{PS} \simeq 0.20$ at the interface. Note that the pinned superconducting amplitude equals the maximum value achievable in the stable uniform bulk as we discuss below. This pinning at the maximum is a central result of the present report. The pinned superconducting order parameter is consistent with the anomalous pinning of $T_c$ independent of $\delta_{\text{bulk}}$ observed at interfaces (*2*). Indeed, it is natural that the same order parameter at $T = 0$ yields the same $T_c$.

A question arises on how robust the results are when the atomic inter-layer diffusion is absent, where the density functional calculation indicates that the onsite level varies relatively suddenly at the interface (Fig. 1 (C)). We show in the Supplementary Materials that the pinning still exists.

**DISCUSSION**

**Relation between intralayer and interlayer phase separations**

In the bulk system, the phase separation and enhancement of charge susceptibility near the Mott insulator were first theoretically pointed out (*21-23*) and has long been debated (*24-33*) in experiments and theories. Even in the simple Hubbard model on the square lattice, the exact solution is not available in terms of the existence of the phase separation and superconductivity (see Ref. (*10*) for detailed comparisons of sometimes controversial theoretical results and their accuracies). Above all, many of accurate numerical results have indicated the existence of an extended region of phase separation region for large $U/t$. Furthermore, it was shown that the superconducting correlation exhibits its maximum inside the phase separation region, if we allow for metastable states (*10*). However, as a thermodynamically stable state, the maximum emerges at the phase separation border $\delta = \delta_{PS}$. The region $0 < \delta < \delta_{PS}$ is subtle: Here the long-ranged Coulomb repulsion ignored in the Hubbard model would lead to a diverging electrostatic energy, if the macroscopic phase separation occurs, which is prohibited in reality. Consequently, the true ground state is replaced by mesoscale inhomogeneous states or long-period charge order to compromise with the long-ranged Coulomb force as was observed (*34*). Even for the Hubbard model in the absence of long-range interactions, stripe-type charge ordering is nearly degenerate with the phase separated state (*10,28,29,35*). Such inhomogeneities introduce pair breaking and suppresses the superconductivity allowing for the maximum superconducting order only at the pinpoint of $\delta \sim \delta_{PS}$. Even when the stripe (charge density wave) is perfectly ordered and clean, this suppression occurs where the superconductivity at the optimum carrier concentration is connected by the Josephson tunneling through the non-optimized density region, which has a similarity to continuous superconductor-insulator transition caused by randomness (*36-38*).

In contrast, around the interface, our result indicates that the charge inhomogeneity is circumvented by transferring holes between the neighboring metallic layers and the interface to avoid the energy cost caused by the intralayer charge inhomogeneity. This transfer violates the charge neutrality of each layer but the electrostatic energy remains small because it corresponds to the formation of a capacitor, where the electric field is confined only within the capacitor. This is a remarkable way of avoiding the harmful electronic inhomogeneity that is unavoidable in the bulk. At the interface, the inhomogeneity is dissolved into an imbalanced density between the neighboring layers, pinned at both ends of the phase separation region, $\delta = \delta_{PS}$ and 0. In fact the stable hole density at the interface at $\delta_{PS}$ ensures the maximum superconducting amplitude ever realized in the bulk, consistently with the pinned $T_c \sim 40$ K (*2*).

We further discuss our intuitive interpretation of why the interlayer phase separation is more stable than a state with intralayer charge inhomogeneities. Although the divergence of the electrostatic energy is avoided even by the introduction of mesoscale inhomogeneities within a layer, the formation of stripes or puddles costs a boundary energy proportional to the length of the domain wall within the layer. In fact, among the energy cost caused by the domain wall formation, $E_D$, there are two contributions $E_{D1}$ and $E_{D2}$ which will be crucially different between the intra- and inter-layer domain-wall formations.

Because it is in the phase separation region, the energy as a function of doping concentration has a double-well structure, whose two minima are realized at the two phase separated densities. Forming a domain wall within a layer costs energy $E_{D1}$ because in the transient region at the domain wall, the charge density crosses through the maximum in the center of this double-well structure. On the other hand, if the domain wall is located between two layers, this energy cost can be largely avoided because the charge density can jump from low to high values thanks to the small $t_z$ and the presence of the intermediate LaO layers.

The other one, $E_{D2}$ is the cost arising from the spatial dependence of the charge density. Since the Coulomb contribution that arises from the long-ranged part to this spatial dependene is material and model dependent, we do not discuss it in detail. A crucial difference between the inter- and intra-layer phase separations for the spatial dependent part of the energies arises from the kinetic energy: The kinetic energy is clearly lost in the presence of the domain wall because of the carrier confinement in the carrier rich region. The kinetic energy is dominated by the intralayer hopping contribution, therefore, the domain wall within the layer costs more kinetic energy than the interlayer domain wall.

Our finding offers possible ways for enhancing and stabilizing the superconducting amplitude by making use of the translational symmetry breaking in the interlayer direction. Controlling the carrier density such as to reach the endpoint of the phase separation in the bulk is the best way to optimize superconductivity in the uniform bulk. However, it requires a careful tuning. At the interface, the situation is much more robust, since the optimal value is automatically reached in a self-organized manner. One can therefore expect easier routes for materials preparation, than the careful tuning needed in the bulk.

Furthermore, elucidating the pinning mechanism provides guidelines for the design of materials and devices with enhanced superconductivity: A likely strategy is to attempt interface engineering. An example is to keep the carrier density even in the metastable region inside the phase separation region. Here, the superconducting amplitude would be even larger than at the endpoint of the phase separation region.

A related future issue is the mechanism in multi-layer superconductors (*39,40*). In fact, the charge inhomogeneity and the resultant suppression of the superconductivity can be avoided by the external breaking of the translational symmetry as in the case of the interface and the multi-layer systems. Since the iron-based and cuprate superconductors are both on the verge of phase separation (*10, 14*), this strategy may universally apply to materials where high-temperature superconductivity is driven by electron correlation effects.

**MATERIALS AND METHODS**

**Numerical Methods**

To analyze the multi-layer Hubbard model in the ground state, we employ the mVMC method. Here, we summarize this method briefly. Details can be found in Refs. (*41*). The variational wave function for the ground state is defined as

$$|\psi\rangle = \mathcal{P}_\mathrm{G}\mathcal{P}_\mathrm{J}\mathcal{L}^S |\phi_\mathrm{pair}\rangle, \quad (4)$$

where $\mathcal{P}_\mathrm{G}$ and $\mathcal{P}_\mathrm{J}$ are the Gutzwiller (*42*) and Jastrow (*43, 44*) factors, respectively (*10, 41*). These correlation factors are defined as

$$\mathcal{P}_\mathrm{G} = \exp(-\sum_{i,\nu} g_\nu\, n_{i\uparrow\nu} n_{i\downarrow\nu}),$$

$$\mathcal{P}_\mathrm{J} = \exp(-\frac{1}{2}\sum_{i,j,\nu,\mu} v_{ij\nu\mu}\, n_{i\nu} n_{j\mu}),$$

where $g_\nu$ and $v_{ij\nu\mu}$ are variational parameters. These factors express many-body correlations beyond the mean-field starting point. To restore the symmetry of the Hamiltonian, we employ the quantum number projection method (*45*). In this study, we use the total spin quantum number projection operator $\mathcal{L}^S$, which restores $SU$(2) spin symmetry with the total spin $S$, where $S = 0$. The one-body part $|\phi_{\text{pair}}\rangle$ is the generalized pairing wave function defined as

$$|\phi_{\text{pair}}\rangle = [\sum_{i,j,\nu,\mu} f_{ij\nu\mu}\, c^\dagger_{i\uparrow\nu} c^\dagger_{j\downarrow\mu}]^{\frac{\mathcal{N}}{2}} |0\rangle, \qquad (5)$$

where $f_{ij\nu\mu}$ denotes the variational parameters and $\mathcal{N}$ represents the total number of electrons. In this study, we allow $f_{ij\nu\mu}$ to have a 2 × 2 sublattice structure for each layer (2 × 2 × $L_{\text{layer}}$ sites exist in the unit cell). We note that the variational wave function $|\psi\rangle$ defined in Eq. (4) can flexibly describe different phases such as the antiferromagnetic, the superconducting, and the correlated paramagnetic phases. This flexibility is necessary to analyze the multi-layer model where the competitions and/or coexistence of different phases appear. Although the number of variational parameters becomes large to allow the flexibility, (in this calculation the number of variational parameters is more than $10^4$) we optimize all the variational parameters simultaneously by using the stochastic reconfiguration method (*39, 46*).

In the actual calculations, we take a $L \times L \times L_{\text{layer}}$ lattice with $L = 10$ and $L_{\text{layer}} = 5$ with antiperiodic-periodic (AP) boundary conditions in each layer and in the direction perpendicular to the layers, open boundary conditions at the two end layers. The system size is sufficiently large even when one wishes to examine the long-range order of the superconductivity: We confirmed the saturation of the superconducting correlation at long distances when the superconductivity emerges. The obtained superconducting correlations at each layer are close to those obtained for the uniform bulk simulation for the same hole density with that layer. The superconducting correlation of the uniform bulk does not appreciably depend on the thickness of the uniformly stacked layers if the thickness exceeds three layers while it is slightly smaller than the single layer result. The small difference originates from the small interlayer hopping $t_z$.

**Method to determine charge profile**

We define the chemical potential of each layer after taking into account many-body effects as

$$\mu_\nu(\overline{N}_\nu) = [E_\nu(\mathcal{N}) - E_\nu(\mathcal{N}')]/(N_\nu(\mathcal{N}) - N_\nu(\mathcal{N}')), \quad (6)$$

$$N_\nu(\mathcal{N}) = \sum_{i,\sigma} \langle c_{i\sigma\nu}^\dagger c_{i\sigma\nu} \rangle, \quad (7)$$

$$E_\nu(\mathcal{N}) = -t \sum_{\langle i,j \rangle,\sigma} \langle c_{i\sigma\nu}^\dagger c_{j\sigma\nu} + h.c. \rangle + U \sum_i \langle n_{i\uparrow\nu} n_{i\downarrow\nu} \rangle, \quad (8)$$

where $\overline{N}_\nu = (N_\nu(\mathcal{N}) + N_\nu(\mathcal{N}'))/2$ and $E_\nu(\mathcal{N})$ ($N_\nu(\mathcal{N})$) denotes the total energy (electron number) at the νth layer, when the total electron number of the multi-layer slab is $\mathcal{N}$. Here, we ignore the negligible contribution from the interlayer kinetic energy as we remark later. $\mathcal{N}'$ should be close to $\mathcal{N}$ to approximate the derivative by the difference in Eq. (6). In the definition of $E_\nu(\mathcal{N})$, the site indices $i$ and $j$ run over the sites contained within the νth layer.

For several choices of Δε, we show $\delta_\nu$ (hole density at the νth layer) dependence of $\mu_\nu$ in Fig. 4(A) for ν= 4, which is obtained by changing the total electron number in the canonical ensemble of the slab. Here, the hole density and the chemical potential in the bulk layer at ν = 4, $\delta_4$ and $\mu_4$, respectively have to satisfy the relation between the bulk hole density ($\delta_{bulk}$) and the bulk chemical potential ($\mu_{bulk}$) calculated independently in the uniform bulk system. For the latter, we use the result of the single layer (*10*) because of the periodicity of the bulk and negligible contribution of $t_z$. We separately confirmed that uniformly stacked layers (slab) coupled by $t_z$=0.05$t$ does not give difference in δ dependence of μ regardless of the layer thickness of the slab. The μ-δ relation is shown as the green solid curve without symbols in Fig. 4(A). This poses a constraint that the total electron number in the canonical ensemble of the slab is uniquely determined when we fix $\Delta\varepsilon$. Namely, the point where the doping dependence of $\mu_4$ crosses with the chemical potential of the bulk ($\mu_{bulk}$) represents the true bulk hole density for a given $\Delta\varepsilon$. For instance, for $\Delta\varepsilon = 0.2$, $\mu_4$ crosses with the $\mu_{bulk}$ around $\delta_4 \sim 0.32$. We then employ $\delta_4 \sim 0.32$ as the bulk hole density $\delta_{bulk}$ for $\Delta\varepsilon = 0.2$. The results shown in the main text are obtained from the calculations that satisfy this constraint. Figures 4(B) and (C) show that the relation between the chemical potential $\mu$ and the hole density $\delta$ in each layer follows the relation for the uniform bulk, confirming that the interlayer transfer does not change this relation.

**REFERENCES AND NOTES**

1. J.-F. Ge, Z.-L. Liu, C. Liu, C.-L. Gao, D. Qian, Q.-K. Xue, Y. Liu, J.-F. Jia, Superconductivity above 100 K in single-layer FeSe films on doped SrTiO$_3$. *Nat. Mater.* **14**, 285-289 (2015).

2. J. Wu, O. Pelleg, G. Logvenov, A. T. Bollinger, Y-J. Sun, G. S. Boebinger, M. Vanević, Z. Radović, I. Božović, Anomalous independence of interface superconductivity from carrier density. *Nat. Mater.* **12**, 877-881 (2013).

3. S. Sorella, G. B. Martins, F. Becca, C. Gazza, L. Capriotti, A. Parola, E. Dagotto, Superconductivity in the Two-Dimensional $t-J$ Model. *Phys. Rev. Lett.* **88**, 117002 (2002).

4. M. Aichhorn, E. Arrigoni, M. Potthoff, W. Hanke, Antiferromagnetic to superconducting phase transition in the hole- and electron-doped Hubbard model at zero temperature. *Phys. Rev. B* **74**, 024508 (2006).

5. B. Edegger, V. N. Muthukumar, C. Gros, Gutzwiller - RVB theory of high-temperature superconductivity: Results from renormalized mean-field theory and variational Monte Carlo calculations *Adv. Phys.* **56**, 927-1033 (2007).

6. S. S. Kancharla, B. Kyung, D. Sénéchal, M. Civelli, M. Capone, G. Kotliar, A.-M. S. Tremblay, Anomalous superconductivity and its competition with antiferromagnetism in doped Mott insulators. *Phys. Rev. B* **77**, 184516 (2008).

7. M. Civelli, Doping-driven evolution of the superconducting state from a doped Mott insulator: Cluster dynamical mean-field theory. *Phys. Rev. B* **79**, 195113 (2009).

8. G. Sordi, K. Haule, A. M. S. Tremblay, Mott physics and first-order transition between two metals in the normal-state phase diagram of the two-dimensional Hubbard model. *Phys. Rev. B* **84**, 075161 (2011).

9. E. Gull, O. Parcollet, A. J. Millis, Superconductivity and the Pseudogap in the Two-Dimensional Hubbard Model. *Phys. Rev. Lett.* **110**, 216405 (2013).

10. T. Misawa, M. Imada, Origin of high-$T_c$ superconductivity in doped Hubbard models and their extensions: Roles of uniform charge fluctuations. *Phys. Rev. B* **90**, 115137 (2014).

11. B.-X. Zheng, G. K.-L. Chan, Ground-state phase diagram of the square lattice Hubbard model from density matrix embedding theory. *Phys. Rev. B* **93**, 035126 (2016).

12. D. Tahara, M. Imada, Variational Monte Carlo Method Combined with Quantum-Number Projection and Multi-Variable Optimization. *J. Phys. Soc. Jpn.* **77**, 114701 (2008).

13. S. A. Kivelson, E. Fradkin, V. J. Emery, Electronic liquid-crystal phases of a doped Mott insulator. *Nature* **393**, 550-553 (1998).

14. T. Misawa, M. Imada, Superconductivity and its mechanism in an *ab initio* model for electron-doped LaFeAsO. *Nat. Commun.* **5**, 5738 (2014).

*15.* D. van der Marel, Interface superconductivity: Pinning the critical temperature. *Nat. Mater* **12**, 875-876 (2013).

16. O. K. Andersen, A. I. Lichitenstein, O. Jepsen, E. Paulsen, LDA energy bands, low-energy hamiltonians, $t'$, $t''$, $t_\perp(\mathbf{k})$ and $J_\perp$. *J. Phys. Chem. Solids* **56**, 1573-1591 (1995).

17. S. Watanabe, M. Imada, Precise Determination of Phase Diagram for Two-Dimensional Hubbard Model with Filling- and Bandwidth-Control Mott Transitions: Grand-Canonical Path-Integral Renormalization Group Approach. *J. Phys. Soc. Jpn.* **73**, 1251-1266 (2004).

18. S. Uchida, T. Ido, H. Takagi, T. Arima, Y. Tokura, S. Tajima, Optical spectra of $La_{2-x}Sr_xCuO_4$ : Effect of carrier doping on the electronic structure of the $CuO_2$ plane. *Phys. Rev. B* **43**, 7942-7954 (1991).

19. N. E. Hussey, M. Abdel-Jawad, A. Carrington, A. P. Mackenzie, L. Balicas, A coherent three-dimensional Fermi surface in a high-transition-temperature superconductor. *Nature* **425**, 814-817 (2003).

20. G. Logvenov, A. Gozar, I. Bozovic, High-Temperature Superconductivity in a Single Copper-Oxygen Plane. *Science* **326**, 699 (2009).

21. N. Furukawa, M. Imada, Two-Dimensional Hubbard Model --- Metal Insulator Transition Studied by Monte Carlo Calculation ---, *J. Phys. Soc. Jpn.* **61,** 3331-3354 (1992).

22. N. Furukawa, M. Imada, Charge Mass Singularity in Two-Dimensional Hubbard Model, *J. Phys. Soc. Jpn.* **62** 2557-2560 (1993).

23. V. J. Emery, S. A. Kivelson, Frustrated electronic phase separation and high-temperature superconductors, *Physica C* **209**, 597-621 (1993).

24. A. Ino, T. Mizokawa, A. Fujimori, K. Tamasaku, H. Eisaki, S. Uchida, T. Kimura, T. Sasagawa, K. Kishio, Chemical Potential Shift in Overdoped and Underdoped $La_{2-x}Sr_xCuO_4$, *Phys. Rev. Lett.* **79**, 2101-2104 (1997).

25. I. Bozovic, G. Logvenov, M. A. J. Verhoeven, P. Caputo, E. Goldobin, T. H. Geballe, No mixing of superconductivity and antiferromagnetism in a high-temperature superconductor, *Nature* **422**, 873-875 (2003).

26. K. M. Lang, V. Madhavan, J. E. Hoffman, E. W. Hudson, H. Eisaki, S. Uchida, J. C. Davis, Imaging the granular structure of high-Tc superconductivity in underdoped $Bi_2Sr_2CaCu_2O_{8+\delta}$ *Nature* **415**, 412-416 (2002).

27. M. Capone and G. Kotliar, Competition between *d*-wave superconductivity and antiferromagnetism in the two-dimensional Hubbard model, Phys. Rev. B **74**, 054513 (2006).

28. M. Aichhorn, E. Arrigoni, M. Potthoff, W. Hanke, Phase separation and competition of superconductivity and magnetism in the two-dimensional Hubbard model: From strong to weak coupling. *Phys. Rev. B* **76**, 224509 (2007).

29. C.-C. Chang, S. Zhang, Spin and Charge Order in the Doped Hubbard Model: Long-Wavelength Collective Model. *Phys. Rev. Lett.* **104**, 116402 (2010).

30. E. Khatami, K. Mikelsons, D. Galanakis, A. Macridin, J. Moreno, R. T. Scalettar, M. Jarrell, Quantum criticality due to incipient phase separation in the two-dimensional Hubbard model *Phys. Rev. B* **81**, 201101(R) (2010).

31. S. Sorella, Linearized auxiliary fields Monte Carlo technique: Efficient sampling of the fermion sign, *Phys. Rev. B* **84**, 241110 (2011).

32. G. Sordi, P. Sémon, K. Haule, A.-M. S. Tremblay, Strong Coupling Superconductivity, Pseudogap, and Mott Transition, *Phys. Rev. Lett.* **108**, 216401 (2012).

33. L. F. Tocchio, H. Lee, H. O. Jeschke, R. Valentí, C. Gros, Mott correlated states in the underdoped two-dimensional Hubbard model, Variational Monte Carlo versus a dynamical cluster approximation, *Phys. Rev. B* **87**, 045111 (2013).

34. J. M. Tranquada, B. J. Sternlleb, J. D. Axe, Y. Nakamura, S. Uchida, Evidence for stripe correlations of spins and holes in copper oxide superconductors. *Nature* **375**, 561-563 (1995).

35. P. Corboz, S. R. White, G. Vidal, M. Troyer, Stripes in the two-dimensional *t-J* model with infinite projected entangled-pair states. *Phys. Rev. B* **84**, 041108 (2011).

36. M. Wallin, E. S. Sorensen, S. M. Girvin,, A. P. Young, Superconductor-insulator transition in two-dimensional dirty boson systems, *Phys. Rev. B* **49**, 12115- (1994).

37. Y. Dubi, Y. Meir, Y. Avishai, Nature of the superconductor–insulator transition in disordered superconductors, *Nature* **449**, 876-880 (2007).

38. Y.Fukuzumi, K. Mizuhashi, K. Takenaka, S. Uchida, Universal Superconductor-Insulator Transition and Tc Depression in Zn-Substituted High-Tc Cuprates in the Underdoped Regime *Phys. Rev. Lett.* **76**, 684-687 (1996)

39. H. Mukuda, S. Shimizu, A. Iyo, Y. Kitaoka, High-Tc Superconductivity and Antiferromagnetism in Multilayered Copper Oxides -A New Paradigm of Superconducting Mechanism-. *J. Phys. Soc. Jpn.* **81**, 011008 (2012).

40. J.-Z. Ma, A. van Roekeghem, P. Richard, Z.-H. Liu, H. Miao, L.-K. Zeng, N. Xu, M. Shi, C. Cao, J.-B. He, G.-F. Chen, Y.-L. Sun, G.-H. Cao, S.-C. Wang, S. Biermann, T. Qian, H. Ding, Correlation-Induced Self-Doping in the Iron-Pnictide Superconductor $Ba_2Ti_2Fe_2As_4O$. *Phys. Rev. Lett.* **113**, 266407 (2014).

41. D. Tahara, M. Imada, Variational Monte Carlo Method Combined with Quantum-Number Projection and Multi-Variable Optimization. *J. Phys. Soc. Jpn.* **77**, 114701 (2008).

42. M. C. Gutzwiller, Effect of Correlation on the Ferromagnetism of Transition Metals. *Phys. Rev. Lett.* **10**, 159-162 (1963).

43. R. Jastrow, Many-Body Problem with Strong Forces. *Phys. Rev.* **98**, 1479-1484 (1955).

44. M. Capello, F. Becca, M. Fabrizio, S. Sorella, E. Tosatti, Variational Description of Mott Insulators. *Phys. Rev. Lett.* **94**, 026406 (2005).

45. T. Mizusaki, M. Imada, Quantum-number projection in the path-integral renormalization group method. *Phys. Rev.* B **69**, 125110 (2004).

46. S. Sorella, Generalized Lanczos algorithm for variational quantum Monte Carlo. *Phys. Rev.* B **64**, 024512 (2001).

47. P. Giannozzi, et al., QUANTUM ESPRESSO: a modular and open-source software project for quantum simulations of materials. J. of Phys. : Condens. Matter 21, 395502 (2009). http://www.quantum-espresso.org/.

48. J. P. Perdew, K. Burke, M. Ernzerhof, Generalized Gradient Approximation Made Simple. Phys. Rev. Lett. 77, 3865-3868 (1996).

49. N. Troullier, J. L. Martins, Efficient pseudopotentials for plane-wave calculations. Phys. Rev. B 43, 1993-2006 (1991).

50. L. Kleinman, D. M. Bylander, Efficacious Form for Model Pseudopotentials. Phys. Rev. Lett. 48, 1425-1428 (1982).

51. S. G. Louie, S. Froyen, M. L. Cohen, Nonlinear ionic pseudopotentials in spin-density-functional calculations. Phys. Rev. B 26, 1738-1742 (1982).

52. I. Souza, N. Marzari, D. Vanderbilt, Maximally localized Wannier functions for entangled energy bands. Phys. Rev. B 65, 035109 (2001).

**Acknowledgements: Funding:** We thank the computational resources of the K computer provided by the RIKEN Advanced Institute for Computational Science through the HPCI System Research project (hp140215, hp150211, hp150173, and hp160201) supported by Ministry of Education, Culture, Sports, Science, and Technology (MEXT) of Japan. This work was also supported by Grant-in-Aid for Scientific Research (16H06345, 16K17746) from the MEXT of Japan. We also thank numerical resources in the ISSP Supercomputer Center at University of Tokyo. This work was further supported by the European Research

Council under its Consolidator Grant scheme (project number 617196) and IDRIS/GENCI Orsay under project t2016091393.

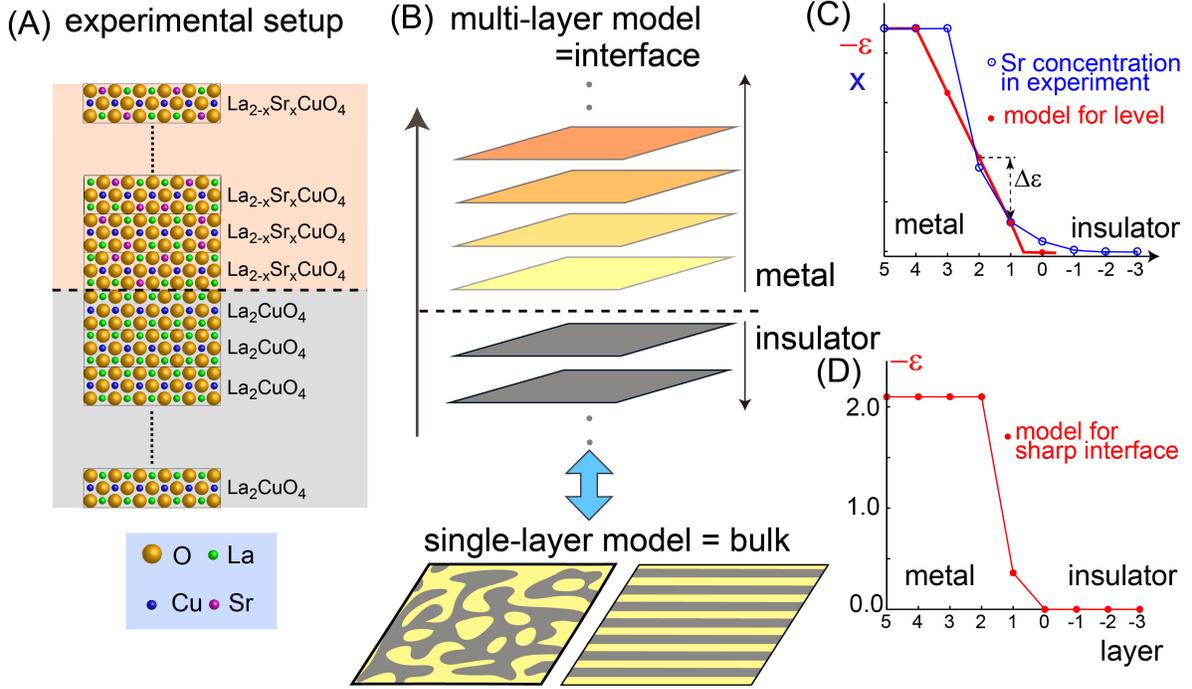

Figure 1: **Experimental setup and present theoretical model of cuprate interface. (A)** Schematic experimental setup of cuprate interface (*2*). **(B)** Top: Illustration of interface model for cuprates. The dotted line denotes the interface between the metallic and insulating layer. The color schematically illustrates the change in the carrier concentration obtained in the present work. Bottom: two hypothetical bulk or single-layer phases with charge inhomogeneity within a layer. **(C)** Layer dependence of onsite level energy chosen to model the interface (red line). In the metallic phase, the onsite energy level is assumed to change linearly. This is an approximation to take into account the effect of interlayer atomic diffusion (blue curve taken from Ref. (*20*)) combined with effects from the Madelung potential and spatial extension of the Wannier orbital at the interface. **(D)** Onsite level of sharp interface modeled by means of an *ab initio* calculation for $x = 0.4$.

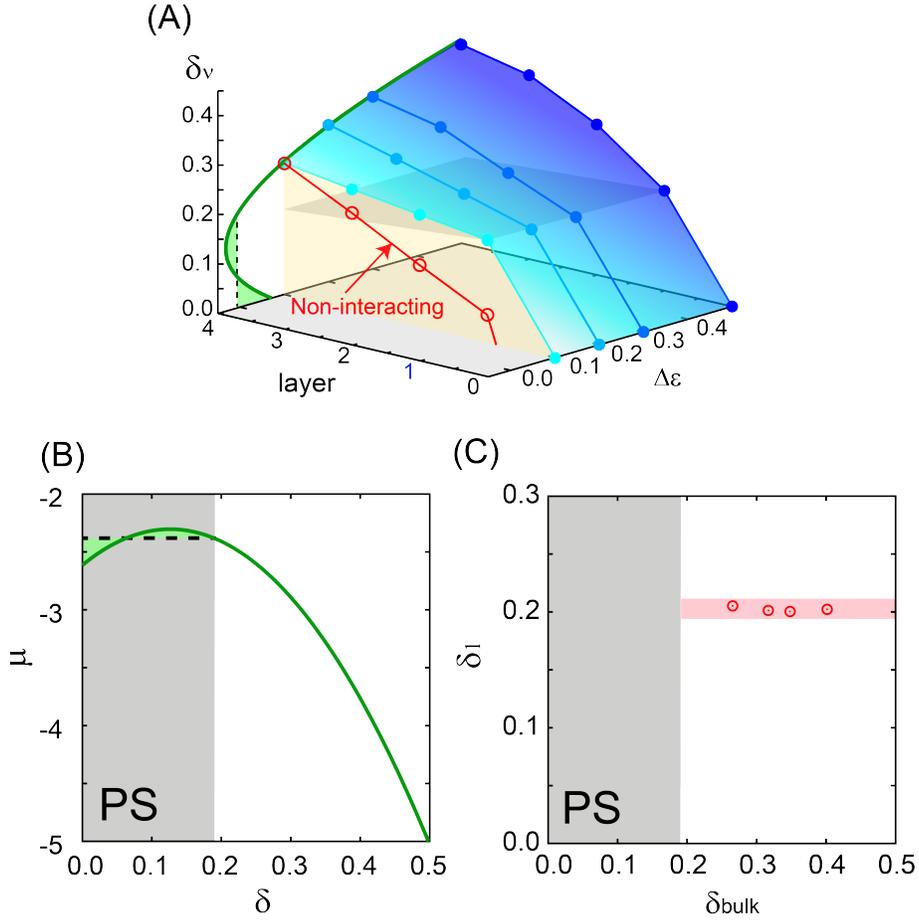

Figure 2: **Layer dependence of doping concentration around interface.** **(A)** Layer and level-slope dependence of carrier density (filled circles and blue surface). At the 4th layer, the green curve is taken from the $\mu_{bulk}$-$\delta_{bulk}$ relation and the two horizontal gray sheets show the phase separation boundaries determined in (B). Note that $\mu_{bulk} = \mu_4 \sim \varepsilon_4 - 2.4$ is satisfied indicating that the grand canonical ensemble is realized for $\nu = 4$. The phase separation region in the bulk is also evaded around the interface in any layer $\nu$. In contrast, the noninteracting case with the same $\delta_4$ plotted for $\Delta\varepsilon = 0.1$ (red line) enters the present phase-separation region. **(B)** Relation between the hole density $\delta_{bulk} = \delta$ and the chemical potential $\mu_{bulk} = \mu$ in the uniform bulk system calculated within the canonical ensemble for a single layer representative of the bulk (*10*). The Maxwell construction (dashed line) determines the phase separation as the gray region between $\delta_{bulk} \sim 0.2$ and 0. **(C)** Hole density at interfaces $\delta_1$ shows pinning against bulk hole density $\delta_{bulk}$.

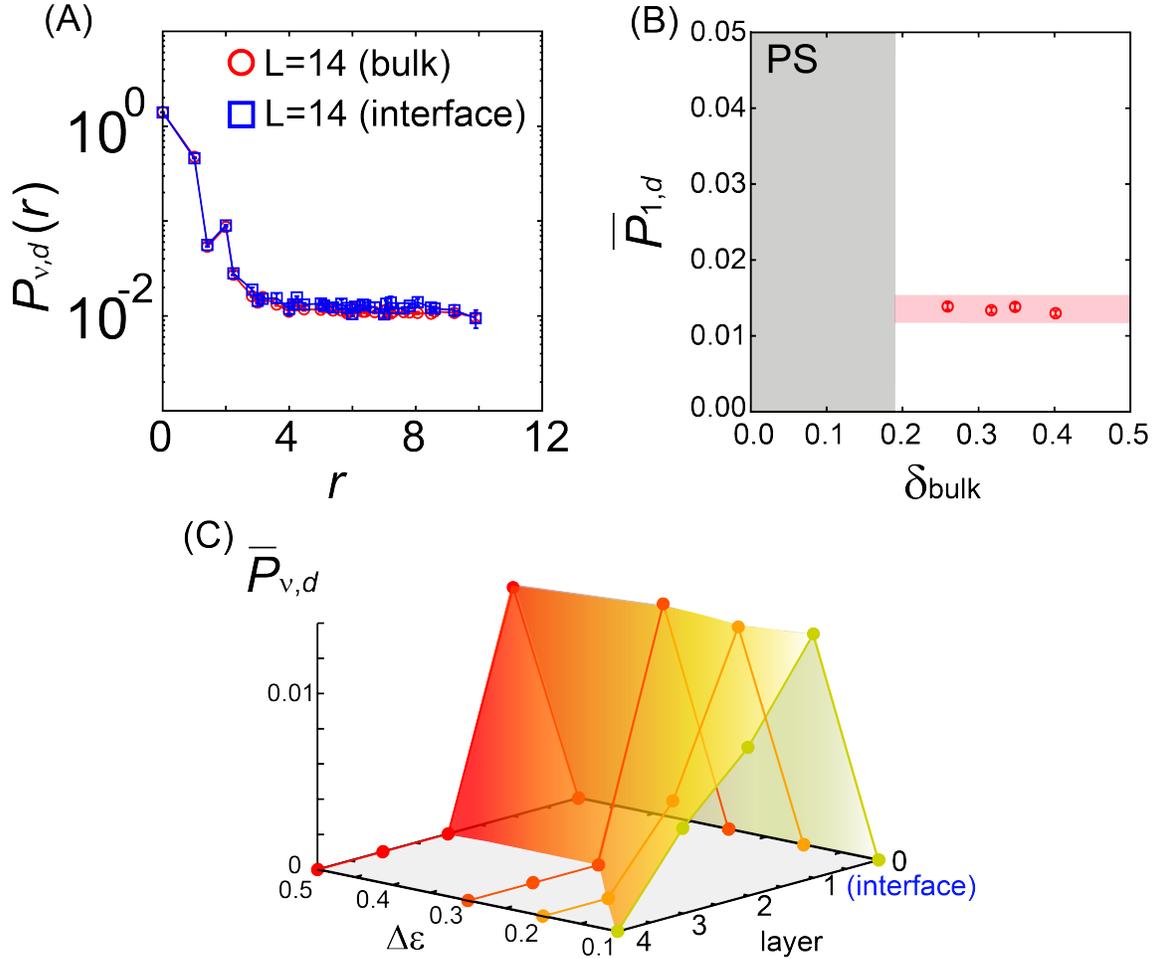

Figure 3: **Superconducting correlations and amplitudes (A)** Spatial dependence of $d$-wave superconducting correlations at the interface ($\nu = 1$) for $\Delta\varepsilon = 0.2$ and $\delta_{bulk} \sim 0.32$ (blue squares) compared with that of the uniform bulk for a hole density similar to that at the interface (~0.20). The red circles are obtained for the bulk (stacked layers) with uniform chemical potential. The saturation at long distances $r$ indicates long-range order. The data sets are both for the linear size in the plane direction, $L = 14$, for which we confirmed

convergence to the thermodynamic limit. **(B)** Bulk hole density dependence of squared superconducting amplitude at the interface ($\nu = 1$) defined by $\overline{P}_{1,d} \cdot \overline{P}_{1,d}$ hardly depends on the bulk hole densities. **(C)** Layer dependence of $\overline{P}_{\nu,d}$. This function is strongly peaked at the interface $\nu = 1$.

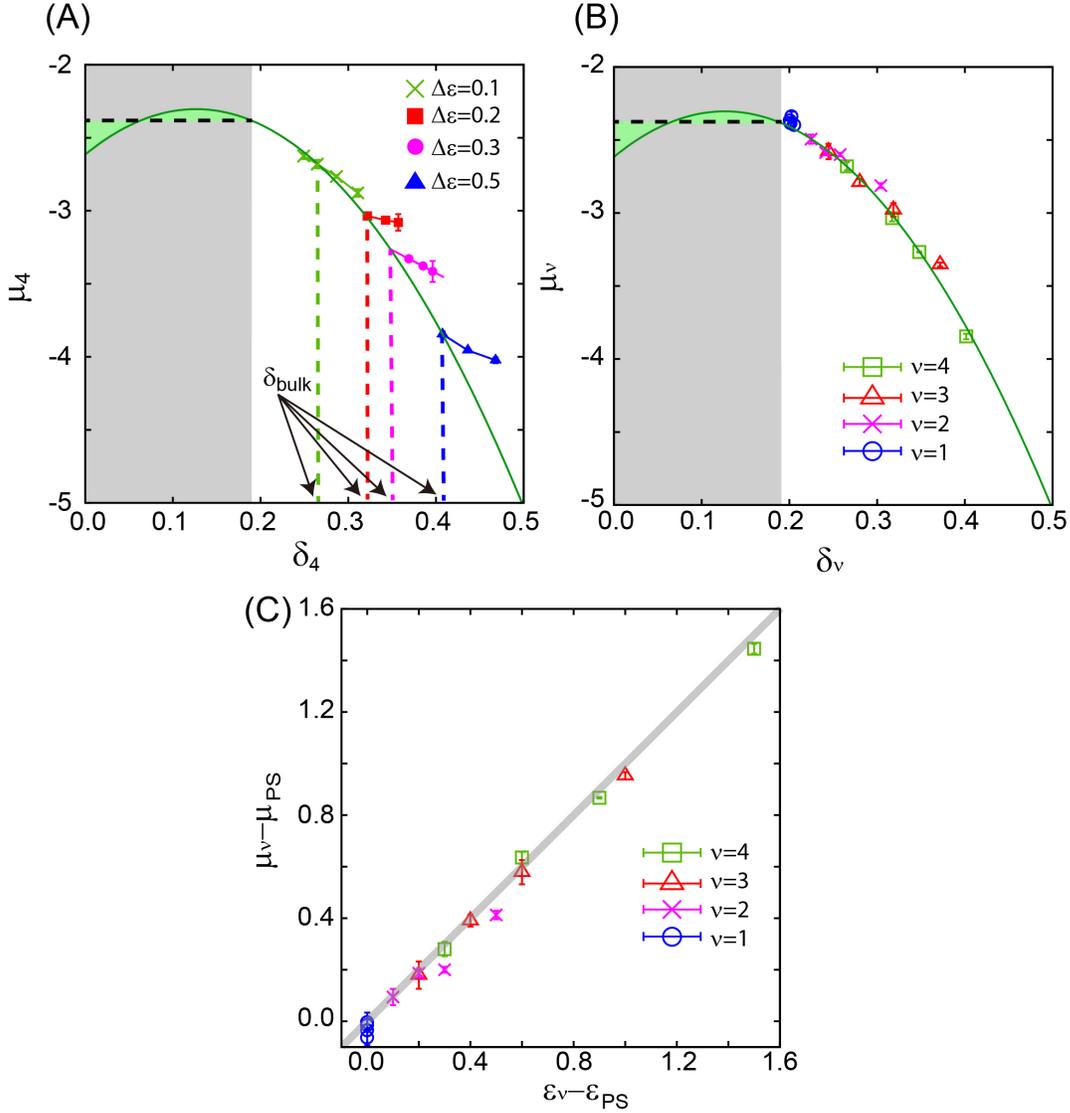

Figure 4: **Relation between chemical potential and hole concentration.** **(A)** Chemical potential $\mu_4$ (determined from Eq. (6)) as a function of the hole density $\delta_4$ at the fourth layer for several choices of $\Delta\varepsilon$ are plotted as curves with symbols. Here, $\delta_\nu$ is defined as $\delta_\nu =$

$1 - \overline{N}_\nu/(L \times L)$. For a choice of $\Delta\varepsilon$, $\mu_4$ curves are drawn by changing the total electron number in the whole slab. We assume that $\mu_4$ converges to the bulk chemical potential $\mu_{bulk}$ (green curve), which is calculated in Ref. (*10*).Therefore, the realistic bulk hole density $\delta_{bulk}$ is determined from the crossing point between the bulk chemical potential $\mu_{bulk}$ given by the green curve and $\mu_4$ for each choice of $\Delta\varepsilon$. Cases with different $\delta_{bulk}$ are obtained from different $\Delta\varepsilon$. The non-monotonic behavior of the green curve signals the existence of a phase separation region. A Maxwell construction (horizontal dashed line) allows us to determine the coexistence region as $0 < \delta < \delta_{PS} \sim 0.2$ (gray area). **(B)** Chemical potential $\mu_\nu$ (determined from Eq. (6)) as a function of the hole density $\delta_\nu$ for $\nu$ = 1 to 4 for several choices of $\Delta\varepsilon$ are plotted as symbols. It follows the bulk behavior shown by the green curve, indicating that each layer behaves as a single layer (or uniform bulk) in the $\mu$-$\delta$ relation with negligible effects from $t_z$. **(C)** Chemical potential difference $\mu_\nu - \mu_{PS}$ plotted as a function of the onsite level difference $\varepsilon_\nu - \varepsilon_{PS}$. The straight bold line shows that the chemical potential at each layer shifts in accordance with the shift of the onsite level indicating again that the effects of $t_z$ is negligible and each layer behaves as grand canonical ensemble with the hole onsite level $\varepsilon_\nu$.

# Supplementary Materials for Self-Optimized Superconductivity Attainable by Interlayer Phase Separation at Cuprate Interfaces

Takahiro Misawa, Yusuke Nomura, Silke Biermann and Masatoshi Imada*
Corresponding author. E-mail imada@ap.t.u-tokyo.ac.jp

**A. First principles estimate of onsite energy level at sharp interface**

We perform band structure calculations within density-functional theory (DFT) for the supercell containing 16 formula units. We employ the QUANTUM ESPRESSO package (*47*). We assume that Sr is doped uniformly into the doped region (layers between 9 and 16) with a doped hole concentration of 40 %, while layers from 1 to 8 consist of undoped $La_2CuO_4$. This means that the interface is assumed to be sharp and we do not assume any diffusion of Sr. The Sr doping is simulated by the virtual crystal approximation. For the exchange-correlation functional, we adopt the generalized-gradient approximation (GGA) by Perdew, Burke and Ernzerhof (*48*). We prepare the Troullier-Martins norm-conserving pseudopotentials (*49*) in the Kleinman-Bylander representation (*50*) for La, Cu and O atoms. Nonlinear core corrections (*51*) are applied to the pseudopotentials of La and Cu atoms. We employ 8×8×1 *k* points and the cutoff energy for the plane wave basis is set to 80Ry. The atomic-like Wannier functions for Cu $3d_{x^2-y^2}$ orbitals are constructed (*52*).

In Fig. S1(A), we show the band structure of nondoped and doped systems. In the example, we display a nominally 40 % doped case. We also include the band structure of a supercell calculation with 16 formula units in the interlayer direction. In (B), we show the layer dependence of the energy levels of $d_{x^2-y^2}$ Wannier orbitals for the supercell. It reveals that the onsite level nearly abruptly shifts at the interface if the interface is sharply constructed without interlayer atom diffusion. In (C), the momentum resolved energy shift of the Kohn-Sham energy level is plotted, which shows an abrupt shift at the interface similarly to (B). Cases with nominal 20, and 30 % doping in the *ab initio* calculation are also calculated to derive the electronic levels. In the actual many-body calculation, we employ the layer dependent onsite level simplified from the level at (π/2, π/2) momentum at each layer shown in Fig. S1(B) combined with a double counting correction $-Un_\nu^{GGA}/2$. Here, $n_\nu^{GGA}$ is the occupation of the νth layer within the GGA calculation calculated using Wannier orbitals constructed from the whole Cu 3*d* and O 2*p* manifold. The reason why we employ the level at (π/2, π/2) is that the Fermi surface evolves there and its level shift by the GGA may represent more reliably the carrier doping contrary to (π,0), where the pseudogap drastically modifies the electronic structure from the GGA prediction. In any case we are not interested in details of the model, but our purpose here is to see the universality of the pinning mechanism by simulating a hypothetical interface in a case opposite to that with the interlayer atomic diffusion.

**B. Superconducting properties of sharp interface stemming from first-principles estimate**

By using the electronic level as modeled in Fig. S1, we have calculated the layer dependence of the hole density and the superconducting correlation by using the mVMC, by modeling the layer dependence of the onsite energy level. An example is given in Fig. 1(D) for a nominal doping concentration of 40% in GGA shown in Fig. S1. Note that the obtained hole density is different from the nominal hole density in the GGA calculation because of many-body effects.

Even for the relatively abrupt jump of the onsite level stemming from the GGA calculation of the sharp interface, the pinning of the hole density as well as the superconductivity is again seen and the pinning mechanism turns out to be robust as we see in Fig. S2. The first principles calculations are helpful in determining the onsite energy level when the Sr concentration is given. In fact, Fig. S1 demonstrates that the onsite level changes very quickly within a layer or two following the change in Sr concentration.

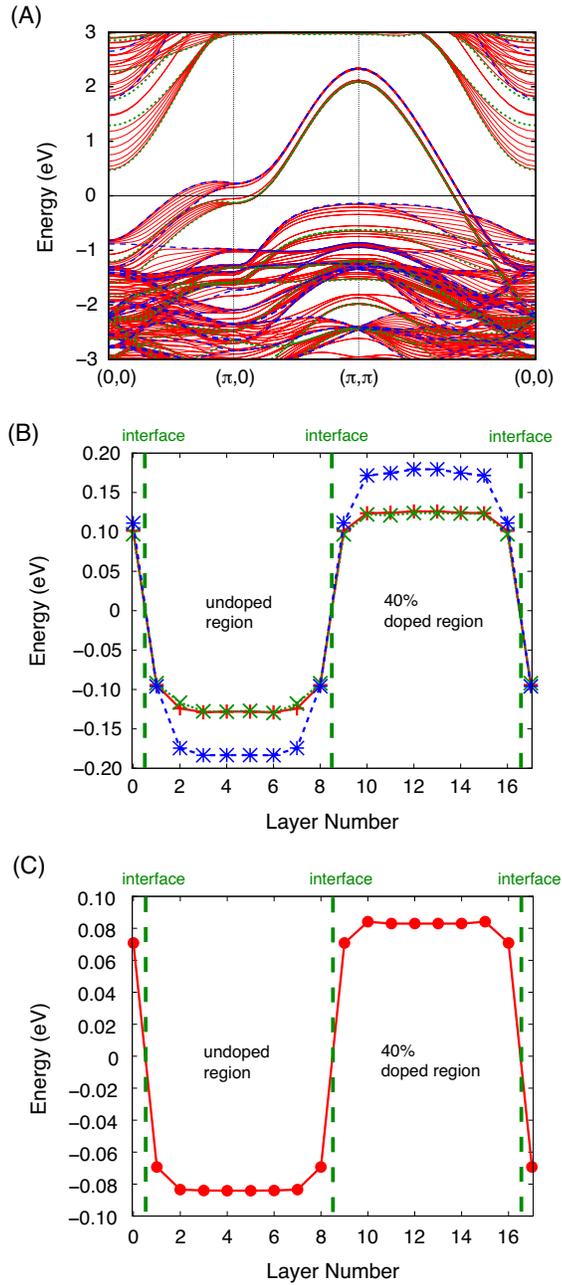

Figure S 1: **First principles estimate of electronic structure around sharp interface.** (A) GGA Band structure of bulk and supercell with 16 formula units at $k_z = 0$. Blue (dark) dashed and green (light) dotted lines are bulk bands for the 40%-doped and undoped systems, respectively. The conventional (tetragonal) cell containing 2 formula units is used

in the calculation. Red solid lines are bands obtained with 16 formula-unit supercell in the interlayer direction and with 8 × 8 × 1 periodic boundary condition. **(B)** Layer dependence of Kohn-Sham (KS) energy levels of $d_{x^2-y^2}$ band at several momenta. We associate each KS level to each layer by using the projection of the KS eigenstates onto the Wannier orbitals of each layer. Blue *, red + and green × are for (π, 0), (π, π) and (π/2, π/2), respectively. The layer that has a maximum weight at a KS eigenstate is used to label this KS state. **(C)** Layer dependence of electronic levels of local $d_{x^2-y^2}$ Wannier orbitals. In (B) and (C), the center of the energy is set to be zero.

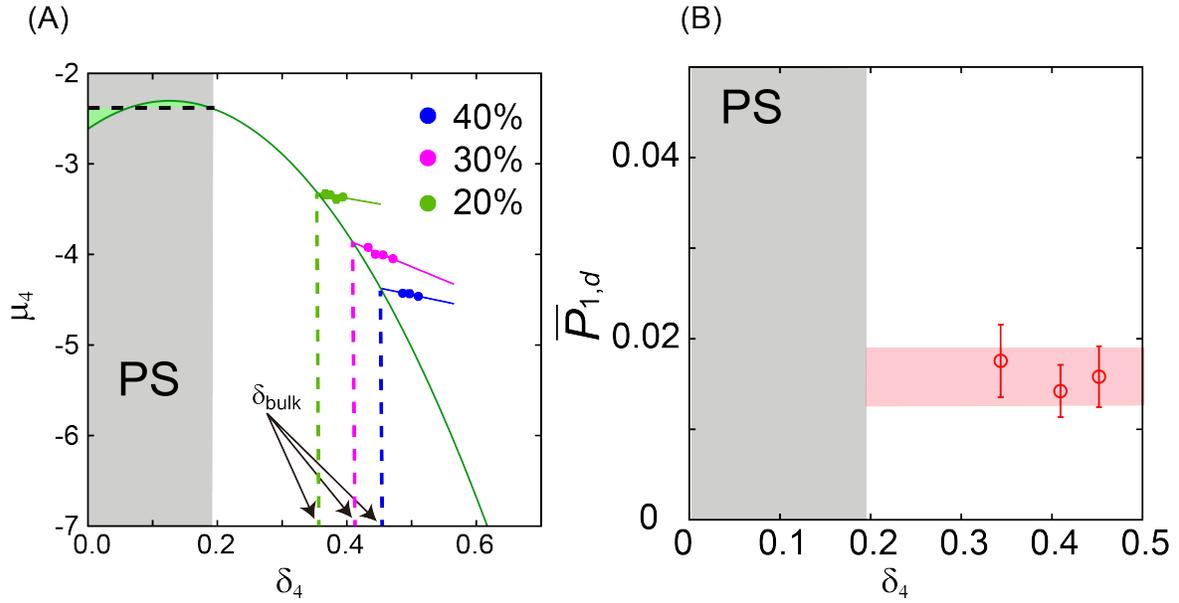

Figure S 2: **Chemical potential in bulk metal and saturated superconducting correlation at interface as functions of bulk hole concentration in case of sharp interface.** (A) Plot obtained in the same way as Fig. 4(A), but for sharp interface illustrated as in Fig. 1(D). Here, the onsite levels are provided from the first-principles calculations with three different doping concentrations (20,30 and 40 %) in the metallic side. (B) Square of superconducting amplitude at long distances $\overline{P}_{1,d}$ at interface. The pinning is similar to the one presented in Fig. 3(B).